\documentclass[
    ,final            
  ,numberedheadings 
  ]
  {aipproc}

\layoutstyle{6x9}

\makeatletter
\renewcommand{\theequation}{\arabic{section}.\arabic{equation}}
\newcommand{\vereq}[2]{\lower3pt\vbox{\baselineskip1.5pt \lineskip1.5pt
\ialign{$\m@th#1\hfill##\hfil$\crcr#2\crcr\sim\crcr}}}

\makeatother

\begin{document}

\title{
\hspace*{12cm}
\hfill{\normalsize\vbox{\hbox{\rm DPNU-03-17}}}\\
\vspace{0.3cm}
Vector Manifestation in Hot Matter
and Violation of Vector Dominance~\footnote{%
Talk given by M.~Harada at MRST 2003 ``Joefest'',
May 13-15, 2003, Syracuse, New York, USA.
}
}

\author{Masayasu Harada}{
  address={Department of Physics, Nagoya University, 
  Nagoya, 464-8602, Japan}
}

\author{Chihiro Sasaki}{
  address={Department of Physics, Nagoya University, 
  Nagoya, 464-8602, Japan}
}

\begin{abstract}
We summarize main mechanisms to realize the
vector manifestation (VM), in which the massless
vector meson becomes the chiral partner of pion, at the
critical temperature in hot QCD within the framework 
of the model based on the hidden local symmetry.
Then, we show a recent analysis 
on the
direct photon-$\pi$-$\pi$ coupling which measures the validity of the
vector dominance (VD) of the electromagnetic form factor of the pion:
The VM predicts that the
VD is largely violated at the critical temperature.
\end{abstract}

\maketitle


\section{I\lowercase{ntroduction}}

Spontaneous chiral symmetry breaking is one of the 
most important properties of QCD in low energy region.
This chiral symmetry is expected to be restored in hot and/or
dense QCD
and properties of hadrons will be changed near the critical temperature
of the chiral symmetry restoration~\cite{rest1,%
Brown-Rho:96,
Rapp-Wambach:00
}.
The CERN Super Proton Synchrotron (SPS) observed
an enhancement of dielectron ($e^+e^-$) mass spectra
below the $\rho / \omega$ resonance~\cite{Agakishiev:1995xb}.
This can be explained by the dropping mass of the $\rho$ meson
(see, e.g., Refs.~\cite{Li:1995qm, Brown-Rho:96, Rapp-Wambach:00})
following the Brown-Rho scaling proposed in Ref.~\cite{BR}.
Furthermore,
the Relativistic Heavy Ion Collider (RHIC) has started
to measure several physical processes in hot matter
which include the dilepton energy spectra.
Therefore it is interesting to study the temperature dependence
of the vector meson mass which is one of the important quantities
in the chiral phase transition.

In Ref.~\cite{HS:VMT},
we showed how the vector manifestation (VM)~\cite{HY:VM}, 
in which the chiral symmetry is restored by 
the massless degenerate pion (and its flavor partners) and the
vector meson (and its flavor partners) as the chiral partner,
can be realized in hot matter using the model for pion and
vector meson based on the
hidden local symmetry (HLS)~\cite{BKUYY}.
There, the {\it intrinsic temperature dependences}~\cite{HS:VMT}
of the parameters of the HLS Lagrangian,
which is introduced by applying
the Wilsonian matching~\cite{HYa} at non-zero temperature,
played important roles
to realize the chiral symmetry restoration consistently:
In the framework of the HLS the equality between 
the axialvector and vector current correlators at critical point
can be satisfied only if the intrinsic thermal effects are
included.

So far, several predictions of the VM in hot matter were made:
The vector and axialvector susceptibilities are predicted to be
equal~\cite{HKRS};
The pion velocity becomes the speed of light when we neglect 
the small Lorentz violating effects in the bare HLS 
Lagrangian~\cite{HKRS};
the vector dominance
of the electromagnetic form factor of pion
is largely violated~\cite{HS:VD}.
Quite recently in Ref.~\cite{Sasaki:Vpi}, the Lorentz breaking
effects were included in the bare HLS Lagrangian, and it was shown 
that the pion velocity at the critical temperatuer receives neither
quantum nor hadronic thermal corrections.

In this write-up we show how the VM is realized in hot QCD
following Refs.~\cite{HS:VMT,HS:VD},
and then
we summarize a main result obtained in 
Ref.~\cite{HS:VD}:
{\it The VM predicts the large violation of the vector
dominance at the critical temperature}.

This write-up is organized as follows:
In section~\ref{sec:VM}, we 
briefly review the difference between the VM and
the conventional picture based on the linear sigma model.
In section~\ref{sec:HLS} we briefly introduce the HLS model.
Section~\ref{sec:ITE} is devoted to review the way to 
incorporate the
intrinsic temperature dependences of the bare HLS parameters
through the Wilsonian matching.
Then, in section~\ref{sec:VMT}, we summarize
how the VM is realized in hot matter.
Section~\ref{sec:PaVVD} is a main part of this write-up
where we show that the VM predicts that
the vector dominance is largely violated 
at the critical temperature.
Finally in section~\ref{sec:Sum}, we 
give a brief summary.

\setcounter{equation}{0}
\section{V\lowercase{ector} M\lowercase{anifestation}}
\label{sec:VM}

In this section, following Ref.~\cite{HYc},
we briefly review the difference between the VM and
the conventional picture based on the linear sigma model
in terms of the chiral representation of the mesons
by extending the analyses done in
Refs.~\cite{Gilman-Harari,Weinberg:69}
for two flavor QCD.

The vector manifestation (VM) was first proposed in
Ref.~\cite{HY:VM} as a novel manifestation of Wigner 
realization of
chiral symmetry where the vector meson $\rho$ becomes massless at the
chiral phase transition point. 
Accordingly, the (longitudinal) $\rho$ becomes the chiral partner of
the Nambu-Goldstone boson $\pi$.
The VM is characterized by
\begin{equation}
\mbox{(VM)} \qquad
F_\pi^2 \rightarrow 0 \ , \quad
m_\rho^2 \rightarrow m_\pi^2 = 0 \ , \quad
F_\rho^2 / F_\pi^2 \rightarrow 1 \ ,
\label{VM def}
\end{equation}
where $F_\rho$ is the decay constant of 
(longitudinal) $\rho$ at $\rho$ on-shell.
This is completely different from 
the conventional picture based
on the linear sigma model 
where the scalar meson becomes massless
degenerate with $\pi$ as the chiral partner:
\begin{equation}
\mbox{(GL)} \qquad
F_\pi^2 \rightarrow 0 \ , \quad
m_S^2 \rightarrow m_\pi^2 = 0 \ .
\label{GL def}
\end{equation}
In Ref.~\cite{HYc}
this was called GL manifestation after the
effective theory of Ginzburg--Landau or Gell-Mann--Levy.

We first consider 
the representations of 
the following zero helicity ($\lambda=0$) states
under
$\mbox{SU(3)}_{\rm L}\times\mbox{SU(3)}_R$;
the $\pi$, the (longitudinal) $\rho$, the (longitudinal) axialvector
meson denoted by $A_1$ ($a_1$ meson and its flavor partners)
and the scalar meson denoted by $S$.
The $\pi$ and the longitudinal $A_1$ 
are admixture of $(8\,,\,1) \oplus(1\,,\,8)$ and 
$(3\,,\,3^*)\oplus(3^*\,,\,3)$
since the symmetry is spontaneously
broken~\cite{Weinberg:69,Gilman-Harari}:
\begin{eqnarray}
\vert \pi\rangle &=&
\vert (3\,,\,3^*)\oplus (3^*\,,\,3) \rangle \sin\psi
+
\vert(8\,,\,1)\oplus (1\,,\,8)\rangle  \cos\psi
\ ,
\nonumber
\\
\vert A_1(\lambda=0)\rangle &=&
\vert (3\,,\,3^*)\oplus (3^*\,,\,3) \rangle \cos\psi 
- \vert(8\,,\,1)\oplus (1\,,\,8)\rangle  \sin\psi
\ ,
\label{mix pi A}
\end{eqnarray}
where the experimental value of the mixing angle $\psi$ is 
given by approximately 
$\psi=\pi/4$~\cite{Weinberg:69,Gilman-Harari}.  
On the other hand, the longitudinal $\rho$
belongs to pure $(8\,,\,1)\oplus (1\,,\,8)$
and the scalar meson to 
pure $(3\,,\,3^*)\oplus (3^*\,,\,3)$:
\begin{eqnarray}
\vert \rho(\lambda=0)\rangle &=&
\vert(8\,,\,1)\oplus (1\,,\,8)\rangle  
\ ,
\nonumber
\\
\vert S\rangle &=&
\vert (3\,,\,3^*)\oplus (3^*\,,\,3) \rangle 
\ .
\label{rhoS}
\end{eqnarray}

When the chiral symmetry is restored at the
phase transition point, 
it is natural to expect that
the chiral representations coincide with the mass eigenstates:
The representation mixing is dissolved.
{}From Eq.~(\ref{mix pi A}) one can easily see~\cite{HY:VM}
that
there are two ways to express the representations in the
Wigner phase of the chiral symmetry:
The conventional GL manifestation
corresponds to 
the limit $\psi \rightarrow \pi/2$ in which
$\pi$ is in the representation
of pure $(3\,,\,3^*)\oplus(3^*\,,\,3)$ 
together with the scalar meson, 
both being the chiral partners:
\begin{eqnarray}
\mbox{(GL)}
\qquad
\left\{
\begin{array}{rcl}
\vert \pi\rangle\,, \vert S\rangle
 &\rightarrow& 
\vert  (3\,,\,3^\ast)\oplus(3^\ast\,,\,3)\rangle\ ,
\\
\vert \rho (\lambda=0) \rangle \,,
\vert A_1(\lambda=0)\rangle  &\rightarrow&
\vert(8\,,\,1) \oplus (1\,,\,8)\rangle\ .
\end{array}\right.
\end{eqnarray}
On the other hand, the VM corresponds 
to the limit $\psi\rightarrow 0$ in which the $A_1$ 
goes to a pure 
$(3\,,\,3^*)\oplus (3^*\,,\,3)$, now degenerate with
the scalar meson in the same representation, 
but not with $\rho$ in 
$(8\,,\,1)\oplus (1\,,\,8)$:
\begin{eqnarray}
\mbox{(VM)}
\qquad
\left\{
\begin{array}{rcl}
\vert \pi\rangle\,, \vert \rho (\lambda=0) \rangle
 &\rightarrow& 
\vert(8\,,\,1) \oplus (1\,,\,8)\rangle\ ,
\\
\vert A_1(\lambda=0)\rangle\,, \vert S\rangle  &\rightarrow&
\vert  (3\,,\,3^\ast)\oplus(3^\ast\,,\,3)\rangle\ .
\end{array}\right.
\end{eqnarray}
Namely, the
degenerate massless $\pi$ and (longitudinal) $\rho$ at the 
phase transition point are
the chiral partners in the
representation of $(8\,,\,1)\oplus (1\,,\,8)$.~\footnote{
  It should be stressed 
  that the VM is realized only as a limit approaching
  the critical point from the broken phase 
  but not exactly on the critical point where the light spectrum
  including the $\pi$ and the $\rho$ would dissappear altogether.
}

Next, we consider the helicity $\lambda=\pm1$. 
Note that
the transverse $\rho$
can belong to the representation different from the one
for the longitudinal $\rho$ ($\lambda=0$) and thus can have the
different chiral partners.
According to the analysis in Ref.~\cite{Gilman-Harari},
the transverse components of $\rho$ ($\lambda=\pm1$)
in the broken phase
belong to almost pure
$(3^*\,,\,3)$ ($\lambda=+1$) and $(3\,,\,3^*)$ ($\lambda=-1$)
with tiny mixing with
$(8\,,\,1)\oplus(1\,,\,8)$.
Then, it is natural to consider in VM that
they become pure $(3\,,\,3^\ast)$ and 
$(3^\ast\,,\,3)$
in the limit approaching the chiral restoration point~\cite{HYc}:
\begin{eqnarray}
\vert \rho(\lambda=+1)\rangle \rightarrow 
  \vert (3^*,3)\rangle\ ,\quad
\vert \rho(\lambda=-1)\rangle \rightarrow 
  \vert (3,3^*)\rangle \ .
\end{eqnarray}
As a result,
the chiral partners of the transverse components of $\rho$ 
in the VM
will be  themselves. Near the critical point the longitudinal $\rho$
becomes 
almost $\sigma$, namely the would-be NG boson $\sigma$ almost 
becomes a 
true NG boson and hence a different particle than the transverse
$\rho$.

\section{H\lowercase{idden} L\lowercase{ocal} S\lowercase{ymmetry}}
\label{sec:HLS}

In this section, we briefly review 
the model based on
the hidden local symmetry (HLS)~\cite{BKUYY}
in which
the vector manifestation is formulated.

The HLS model is based on 
the $G_{\rm{global}} \times H_{\rm{local}}$ symmetry,
where $G=SU(N_f)_L \times SU(N_f)_R$ is the chiral symmetry
and $H=SU(N_f)_V$ is the HLS. 
The basic quantities are 
the HLS gauge boson $V_\mu$ and two matrix valued
variables $\xi_L(x)$ and $\xi_R(x)$
which transform as
 \begin{equation}
  \xi_{L,R}(x) \to \xi^{\prime}_{L,R}(x)
  =h(x)\xi_{L,R}(x)g^{\dagger}_{L,R}\ ,
 \end{equation}
where $h(x)\in H_{\rm{local}}\ \mbox{and}\ g_{L,R}\in
[\mbox{SU}(N_f)_{\rm L,R}]_{\rm{global}}$.
These variables are parameterized as
 \begin{equation}
  \xi_{L,R}(x)=e^{i\sigma (x)/{F_\sigma}}e^{\mp i\pi (x)/{F_\pi}}\ ,
 \end{equation}
where $\pi = \pi^a T_a$ denotes the pseudoscalar Nambu-Goldstone bosons
associated with the spontaneous symmetry breaking of
$G_{\rm{global}}$ chiral symmetry, 
and $\sigma = \sigma^a T_a$ denotes
the Nambu-Goldstone bosons associated with 
the spontaneous breaking of $H_{\rm{local}}$.
This $\sigma$ is absorbed into the HLS gauge 
boson through the Higgs mechanism. 
$F_\pi \ \mbox{and}\ F_\sigma$ are the decay constants
of the associated particles.
The phenomenologically important parameter $a$ is defined as 
 \begin{equation}
  a = \frac{{F_\sigma}^2}{{F_\pi}^2}\ .
 \end{equation}
The covariant derivatives of $\xi_{L,R}$ are given by
\begin{eqnarray}
 D_\mu \xi_L &=& \partial_\mu\xi_L - iV_\mu \xi_L + i\xi_L{\cal{L}}_\mu,
 \nonumber\\
 D_\mu \xi_R &=& \partial_\mu\xi_R - iV_\mu \xi_R + i\xi_R{\cal{R}}_\mu,
\end{eqnarray}
where $V_\mu$ is the gauge field of $H_{\rm{local}}$, and
${\cal{L}}_\mu \ \mbox{and}\ {\cal{R}}_\mu$ are the external
gauge fields introduced by gauging $G_{\rm{global}}$ symmetry.

The HLS Lagrangian with lowest derivative terms at the chiral limit
is given by~\cite{BKUYY}
 \begin{equation}
  {\cal{L}}_{(2)} = {F_\pi}^2\mbox{tr}\bigl[ \hat{\alpha}_{\perp\mu}
                                      \hat{\alpha}_{\perp}^{\mu}
                                   \bigr] +
       {F_\sigma}^2\mbox{tr}\bigl[ \hat{\alpha}_{\parallel\mu}
                  \hat{\alpha}_{\parallel}^{\mu}
                  \bigr] -
        \frac{1}{2g^2}\mbox{tr}\bigl[ V_{\mu\nu}V^{\mu\nu}
                   \bigr]
\ , \label{eq:L(2)}
 \end{equation}
where $g$ is the HLS gauge coupling,
$V_{\mu\nu}$ is the field strength
of $V_\mu$ and
 \begin{eqnarray}
  \hat{\alpha}_{\perp }^{\mu}
     &=& \frac{1}{2i}\bigl[ D^\mu\xi_R \cdot \xi_R^{\dagger} -
                          D^\mu\xi_L \cdot \xi_L^{\dagger}
                   \bigr] \ ,
\nonumber\\
  \hat{\alpha}_{\parallel}^{\mu}
     &=& \frac{1}{2i}\bigl[ D^\mu\xi_R \cdot \xi_R^{\dagger}+
                          D^\mu\xi_L \cdot \xi_L^{\dagger}
                   \bigr]
\ .
 \end{eqnarray}

\setcounter{equation}{0}
\section{I\lowercase{ntrinsic} T\lowercase{hermal} E\lowercase{ffects}}
\label{sec:ITE}

In this section
we briefly review how to extend the Wilsonian matching
to the version at non-zero temperature in order to incorporate
the intrinsic thermal effect into the bare parameters of the HLS
Lagrangian.

We extend the Wilsonian matching proposed at 
$T=0$~\cite{HYa}
to the one at non-zero temperature.
It should be noticed that
there is no longer Lorentz symmetry
in hot matter,
and
the Lorentz non-scalar operators such as
$\bar{q}\gamma_\mu D_\nu q$ may exist in 
the form of the current correlators derived by the 
OPE~\cite{HKL}.
This leads to, e.g., a difference between the temporal and spatial
bare pion decay constants.
However, we neglect the contributions from these operators
since they give a small correction compared with 
the main term $1 + \frac{\alpha_s}{\pi}$.
This implies that the Lorentz symmetry breaking effect in
the bare pion decay constant is small, 
$F_{\pi,\rm{bare}}^t \simeq F_{\pi,\rm{bare}}^s$~\cite{HKRS}.
Thus it is a good approximation that we determine the pion decay
constant at non-zero temperature through the matching 
condition obtained at $T=0$ in Ref.~\cite{HYa}
with putting possible temperature dependences on the gluonic 
and quark condensates~\cite{HS:VMT,HKRS}:
\begin{equation}
 \frac{F^2_\pi (\Lambda ;T)}{{\Lambda}^2} 
  = \frac{1}{8{\pi}^2}\Bigl[ 1 + \frac{\alpha _s}{\pi} +
     \frac{2{\pi}^2}{3}\frac{\langle \frac{\alpha _s}{\pi}
      G_{\mu \nu}G^{\mu \nu} \rangle_T }{{\Lambda}^4} +
     {\pi}^3 \frac{1408}{27}\frac{\alpha _s{\langle \bar{q}q
      \rangle }^2_T}{{\Lambda}^6} \Bigr]
\ .
\label{eq:WMC A}
\end{equation}
Through this condition
the temperature dependences of the quark and gluonic condensates
determine the intrinsic temperature dependences 
of the bare parameter $F_\pi(\Lambda;T)$,
which is then converted into 
those of the on-shell parameter $F_\pi(\mu=0;T)$ 
through the Wilsonian RGEs.

Now, let us consider the Wilsonian matching near the
chiral symmetry restoration point
with assuming that the quark condensate becomes zero
continuously for $T \to T_c$.
First, note that
the Wilsonian matching condition~(\ref{eq:WMC A}) 
provides
\begin{equation}
  \frac{F^2_\pi (\Lambda ;T_c)}{{\Lambda}^2} 
  = \frac{1}{8{\pi}^2}\Bigl[
                            1 + \frac{\alpha _s}{\pi} +
                             \frac{2{\pi}^2}{3}
                            \frac{\langle \frac{\alpha _s}{\pi}
                            G_{\mu \nu}G^{\mu \nu} \rangle_{T_c} }
                             {{\Lambda}^4}
                 \Bigr]
 \neq 0 
\ ,
\label{eq:WMC A Tc}
\end{equation}
which implies that the matching with QCD dictates
\begin{equation}
F^2_\pi (\Lambda ;T_c) \neq 0 
\label{Fp2 Lam Tc}
\end{equation}
even at the critical temperature where the on-shell pion decay
constant 
vanishes by adding the quantum corrections through
the RGE including the quadratic divergence~\cite{HYa}
and hadronic thermal corrections~\cite{HS:VMT,HS:VD}.
As was shown in Ref.~\cite{HKR} for the VM in dense matter,
Lorentz non-invariant version of
the VM conditions for the bare parameters are obtained 
by the requirement of the equality between the axialvector
and vector current correlators in the HLS,
which should be valid also in hot matter~\cite{HKRS}:
\begin{eqnarray}
 && a_{\rm{bare}}^t \equiv
  \Biggl( \frac{F_{\sigma,\rm{bare}}^t}{F_{\pi,\rm{bare}}^t} \Biggr)^2
  \stackrel{T \to T_c}{\to} 1, \quad
  a_{\rm{bare}}^s \equiv
  \Biggl( \frac{F_{\sigma,\rm{bare}}^s}{F_{\pi,\rm{bare}}^s} \Biggr)^2
  \stackrel{T \to T_c}{\to} 1, \label{EVM a}\\
 && g_{T,\rm{bare}} \stackrel{T \to T_c}{\to} 0, \quad
    g_{L,\rm{bare}} \stackrel{T \to T_c}{\to} 0, \label{EVM g}
\end{eqnarray}
where $a^t_{\rm{bare}}, a^s_{\rm{bare}}, g_{T,\rm{bare}}$ and 
$g_{L,\rm{bare}}$ are the extensions of the parameters
$a_{\rm{bare}}$ and $g_{\rm{bare}}$ in the bare Lagrangian
with the Lorentz symmetry breaking effect included as in Appendix A
of Ref.~\cite{HKR}.

When we use the conditions for the parameters $a^{t,s}$ 
in Eq.~(\ref{EVM a})
and the above result that the Lorentz symmetry violation 
between the bare pion decay constants 
$F_{\pi,\rm{bare}}^{t,s}$ is small, 
we can easily show
that the Lorentz symmetry breaking effect between
the temporal and spatial bare $\sigma$ decay constants is also small,
$F_{\sigma,\rm{bare}}^t \simeq F_{\sigma,\rm{bare}}^s$~\cite{HKRS}.
While we cannot determine the ratio $g_{L,\rm{bare}}/g_{T,\rm{bare}}$
through the Wilsonian matching
since the transverse mode of vector meson decouples near
the critical temperature.
However this implies that the transverse mode is irrelevant
to the quantities studied in this paper.
Therefore in the present analysis, we set
$g_{L,\rm{bare}}=g_{T,\rm{bare}}$ for simplicity and
use the Lorentz invariant Lagrangian at bare level.
In the low temperature region, the intrinsic temperature dependences
are negligible, so that we also use the 
Lorentz invariant Lagrangian at bare level
as in the analysis by the ordinary chiral
Lagrangian in Ref.~\cite{GL}.

\setcounter{equation}{0}
\section{V\lowercase{ector} M\lowercase{anifestation in} 
  H\lowercase{ot} M\lowercase{atter}}
\label{sec:VMT}

In this section,
we briefly summarize
how the vector manifestation (VM) is realized in
hot matter following Refs.~\cite{HS:VMT,HS:VD}.

As we discussed in the previous section, 
we start from the Lorentz invariant
bare Lagrangian even in hot matter, and then
the axialvector and the vector
current correlators $G_A^{\rm{(HLS)}}$ and $G_V^{\rm{(HLS)}}$
are expressed by the same forms
as those at zero temperature with the bare parameters
having the intrinsic temperature dependences~\cite{HS:VMT}:
\begin{eqnarray}
 G^{\rm{(HLS)}}_A (Q^2) 
  &=& \frac{F^2_\pi (\Lambda;T)}{Q^2} -
      2z_2(\Lambda;T), \nonumber\\
 G^{\rm{(HLS)}}_V (Q^2) 
  &=& \frac{F^2_\sigma (\Lambda;T)
         [1 - 2g^2(\Lambda;T)z_3(\Lambda;T)]}
           {{M_\rho}^2(\Lambda;T) + Q^2} - 2z_1(\Lambda;T)\ .
  \label{correlator HLS at zero-T}
  \end{eqnarray}

At the critical temperature,
the axialvector and vector current correlators
derived in the OPE
agree with each other for any value of $Q^2$.
Thus we require that
these current correlators in the HLS are
equal at the critical temperature
for any value of $Q^2\ \mbox{around}\ {\Lambda}^2$.
By taking account of the fact 
$F^2_\pi (\Lambda ;T_c) \neq 0$ derived from
the Wilsonian matching condition 
given in Eq.~(\ref{eq:WMC A Tc}),
the requirement 
$G_A^{(\rm{HLS})}=G_V^{(\rm{HLS})}$ is satisfied
only if the following conditions are met~\cite{HS:VMT}: 
\begin{eqnarray}
&&
g(\Lambda;T) \mathop{\longrightarrow}_{T \rightarrow T_c} 0 \ ,
\qquad
a(\Lambda;T) \mathop{\longrightarrow}_{T \rightarrow T_c} 1 \ ,
\nonumber\\
&&
z_1(\Lambda;T) - z_2(\Lambda;T) 
\mathop{\longrightarrow}_{T \rightarrow T_c} 0 \ .
\label{g a z12:VMT}
\end{eqnarray}
These conditions (``VM conditions in hot matter'')
for the bare parameters
are converted into the
conditions for the on-shell parameters through the Wilsonian RGEs.
Since $g=0$ and $a=1$ are separately the fixed points of the RGEs for
$g$ and $a$~\cite{HY:conformal},
the on-shell parameters also satisfy
$(g,a)=(0,1)$, and thus the parametric $\rho$ mass
satisfies $M_\rho = 0$.

Now, let us
include the hadronic thermal effects to obtain the $\rho$ pole
mass near the critical temperature.
As we explained above,
the intrinsic temperature dependences imply that
$M_\rho/T \rightarrow 0$
for $T \rightarrow T_c$,
so that the $\rho$ pole mass near the
critical temperature is expressed as~\cite{HS:VMT,HS:VD}
\begin{eqnarray}
&& m_\rho^2(T)
  = M_\rho^2 +
  g^2 N_f \, \frac{15 - a^2}{144} \,T^2
\ .
\label{mrho at T 2}
\end{eqnarray}
Since $a \simeq 1$ near the restoration point,
the second term is positive. 
Then the $\rho$ pole mass $m_\rho$
is bigger than the parametric
$M_\rho$ due to the hadronic thermal corrections.
Nevertheless, 
{\it the intrinsic temperature dependence determined by the
Wilsonian matching requires
that the $\rho$ becomes massless at the
critical temperature}:
\begin{eqnarray}
&&
m_\rho^2(T)
\rightarrow 0 \ \ \mbox{for} \ T \rightarrow T_c \ ,
\end{eqnarray}
since the first term in Eq.~(\ref{mrho at T 2})
vanishes as $M_\rho\rightarrow 0$, and the second
term also vanishes since $g\rightarrow 0$ for $T \rightarrow T_c$.
This implies that
{\it the vector manifestation (VM) actually
occurs at the critical
temperature}~\cite{HS:VMT}.

\setcounter{equation}{0}
\section{P\lowercase{arameter} $\lowercase{a}$ 
  \lowercase{and} V\lowercase{iolation} \lowercase{of}
   V\lowercase{ector} D\lowercase{ominance} }
\label{sec:PaVVD}

In this section we study the validity 
of vector dominance (VD) of electromagnetic form factor of the pion
in hot matter.
In Ref.~\cite{HY:fate}
it has been shown that VD is 
accidentally satisfied in $N_f=3$ QCD at zero temperature and zero
density, and that it is largely 
violated in large $N_f$ QCD when the VM occurs.
At non-zero temperature there exists the hadronic thermal correction
to the parameters.
Thus it is nontrivial whether or not the
VD is realized in hot matter,
especially near the critical temperature.
Here we will show that the intrinsic temperature dependences
of the parameters
of the HLS Lagrangian play essential roles, and then
the VD is largely violated near the critical temperature.

\subsection{Parameter $a$ at $T=0$}

We first study the direct $\gamma\pi\pi$ interaction at zero
temperature. 
At the leading order of the derivative expansion in the HLS,
the form of the direct $\gamma\pi\pi$ interaction is given by
\begin{equation}
 \Gamma_{\gamma\pi\pi{\rm(tree)}}^\mu 
  = e(q - k)^\mu(1-\frac{a}{2})\ ,
            \label{eq:gpp}
\end{equation}
where 
$e$ is the electromagnetic coupling constant and
$q$ and $k$ denote outgoing momenta of the pions. 
For $a=2$
the direct $\gamma\pi\pi$ coupling vanishes, which leads to
the vector dominance of the electromagnetic form factor of the 
pion~\cite{BKUYY}.

At the next order there exist quantum corrections
which we calculated in the background field gauge in 
Ref.~\cite{HS:VD}.
The resultant direct $\gamma\pi\pi$ interaction in 
the low-energy limit is read 
from the two-point functions of
${\mathcal V}_\mu$-${\mathcal V}_\nu$ and 
${\mathcal A}_\mu$-${\mathcal A}_\nu$:
\begin{equation}
\Gamma_{\gamma\pi\pi}^\mu
=  e \frac{1}{F_\pi^2(0)}
  \left[ 
    q_\nu \Pi_\perp^{\mu\nu}(q) - k_\nu \Pi_\perp^{\mu\nu}(k)
    - \frac{1}{2} (q - k)_\nu \Pi_\parallel^{\mu\nu}(p)
  \right]
\ ,
\label{gpp T0 0}
\end{equation}
where $q$ and $k$ denote the momenta of outgoing pions and
$p_\nu = (q+k)_\nu$ is the photon momentum.

Substituting the decomposition of the two-point function given by
\begin{equation}
 \Pi_{\perp,\parallel}^{\mu\nu}(p)=
  \Pi_{\perp,\parallel}^S(p^2)g^{\mu\nu} +
  \Pi_{\perp,\parallel}^{T}(p^2)(g^{\mu\nu}p^2 - p^\mu p^\nu)\ ,
\label{decomp T0}
\end{equation}
and taking the low-energy limit
$q^2=k^2=p^2=0$, we obtain~\footnote{%
  Note that we adopt 
  $\Pi_{\perp}^S(q^2=0) = F_\pi^2(0)$
  as the renormalization condition for 
  the parametric $F_\pi^2$.
}
\begin{equation}
\Gamma_{\gamma\pi\pi}^\mu
=  e (q-k)^\mu 
  \left[ 
    1
    - \frac{1}{2} \frac{\Pi_\parallel^S(p^2=0)}{F_\pi^2(0)}
  \right]
\ ,
\label{gpp T0}
\end{equation}
where we used $\Pi_\perp^S(q^2=0) = F_\pi^2(0)$.
Comparing the above expression with the one in Eq.~(\ref{eq:gpp}),
we define the parameter $a(0)$ at one-loop level as
\begin{equation}
 a(0) = \frac{\Pi_\parallel^S (p^2=0)}
          {F_\pi^2(0)}\ . \label{eq:Defa}
\end{equation}
We note that, in Ref.~\cite{HYa},
$a(0)$ is defined by the ratio $F_\sigma^2(M_\rho)/F_\pi^2(0)$
with neglecting the finite renormalization effect
which depends on the details of the renormalization condition.
Here we adopt 
\begin{equation}
 \mbox{Re}\Bigl[\Pi_V^S (p^2=M_\rho^2)\Bigr]
  = F_\sigma^2 (\mu = M_\rho), 
\label{Fs ren cond}
\end{equation}
as the renormalization condition for the parametric
$F_\sigma^2$.
{}From this the parameter $a(0)$ is expressed as~\cite{HS:VD}
\begin{equation}
  a(0) =
  \frac{F_\sigma^2(M_\rho)}{ F_\pi^2(0) }
  {}+\frac{N_f}{(4\pi)^2}\frac{M_\rho^2}{F_\pi^2(0)}
     \bigl( 2 - \sqrt{3}\tan^{-1}\sqrt{3} \bigr) 
\ .
\label{a0 expression}
\end{equation}
Using $M_\rho=771.1\,\mbox{MeV}$,
$F_\pi(\mu = 0)=86.4\,\mbox{MeV}$ estimated in the chiral 
limit~\cite{Gas:84} and
$F_\sigma^2(M_\rho)/F_\pi^2(0) = 2.03$ predicted
by the Wilsonian matching 
for $\Lambda_{\rm QCD}=400\,\mbox{MeV}$ and 
the matching scale $\Lambda=1.1\,\mbox{GeV}$
in Ref.~\cite{HYc},
we estimate the value of $a(0)$ at zero temperature 
as~\cite{HS:VD}
\begin{equation}
 a(0) \simeq 2.31 \ .
\label{a0 val}
\end{equation}
This implies that the VD is well satisfied at $T=0$ even though
the value of the parameter $a$ at the scale $M_\rho$ is 
close to one.
It should be noticed that the above result is the predection
of the Wilsonian matching.

\subsection{Parameter $a$ for $T \rightarrow T_c$ and violation of VD}

Now, let us study
the direct $\gamma\pi\pi$ coupling in hot matter.
In general,
the electric mode and the magnetic mode of the photon
couple to the pions differently in hot matter,
so that there are two parameters as extensions of the parameter $a$.
Furthermore, there are four polarization tensors to decompose the
${\mathcal A}_\mu$-${\mathcal A}_\nu$ and
${\mathcal V}_\mu$-${\mathcal V}_\nu$ two-point
functions.
Here we adopt
\begin{eqnarray}
 \Pi_{\perp,\parallel}^{\mu\nu}
&=&
u^\mu u^\nu \Pi_{\perp,\parallel}^t +
   (g^{\mu\nu}-u^\mu u^\nu)\Pi_{\perp,\parallel}^s +
   P_L^{\mu\nu}\Pi_{\perp,\parallel}^L + 
   P_T^{\mu\nu}\Pi_{\perp,\parallel}^T \ ,
\end{eqnarray}
where $u^\mu = (1,\vec{0})$, and $P_L^{\mu\nu}$ and $P_T^{\mu\nu}$
are the polarization tensors.
Similarly to the one obtained at $T=0$ in Eq.~(\ref{gpp T0 0}),
at low-energy limit
the direct $\gamma\pi\pi$ interaction derived from
${\mathcal A}_\mu$-${\mathcal A}_\nu$ and
${\mathcal V}_\mu$-${\mathcal V}_\nu$ two-point
functions is expressed as
\begin{eqnarray}
&&
\Gamma_{\gamma\pi\pi}^{\mu}(p;q,k)
=
\frac{1}{\widetilde{F}(\bar{q};T)
         \widetilde{F}(\bar{k};T)}
\Biggl[
  q_\nu \, 
  \Pi_\perp^{\mu\nu}(q_0,\vec{q};T)
  -
  k_\nu \, 
  \Pi_\perp^{\mu\nu}(k_0,\vec{k};T)
\nonumber\\
&& \qquad\qquad\qquad\qquad\qquad
  {}- \frac{1}{2} (q-k)_\nu \,
    \Pi_\parallel^{\mu\nu}(p_0,\vec{p};T)
\Biggr]
\ ,
\label{direct g pi pi 0}
\end{eqnarray}
where $\bar{q} = \vert\vec{q}\vert$ and
$\bar{k} = \vert\vec{k}\vert$.
$\widetilde{F}$ is the wave function renormalization of 
the $\pi$ field
given by~\cite{HKRS}
\begin{eqnarray}
 \tilde{F}^2(\bar{p};T)
 &=& F_\pi^2(0) + 
  \mbox{Re}\overline{\Pi}_\perp^t(\bar{p},\bar{p};T) \ ,
\label{pi wave renorm}
\end{eqnarray}
where $\overline{\Pi}_\perp^t$ expresses the hadronic thermal 
correction.
In Eq.~(\ref{direct g pi pi 0})
each pion is on its mass shell, so that
$q_0 = v_\pi(\bar{q}) \bar{q}$ and 
$k_0 = v_\pi(\bar{k}) \bar{k}$.
To define extensions of the parameter $a$, 
we consider the soft limit of the photon:
$p_0 \to 0$ and $\bar{p} \to 0$.
Then the pion momenta become
$q_0 = - k_0$ and $\bar{q}=- \bar{k}$.
Note that while
only two components $\Pi_\perp^t$ and $\Pi_\perp^s$ appear
in $q_\nu \, \Pi_\perp^{\mu\nu}$
or $k_\nu \, \Pi_\perp^{\mu\nu}$,
$(q-k)_\nu \, \Pi_\parallel^{\mu\nu}$
includes all four components $\Pi_\parallel^t$,
$\Pi_\parallel^s$, $\Pi_\parallel^L$ and $\Pi_\parallel^T$.
Since the tree part of $\Pi_\parallel^L$ and $\Pi_\parallel^T$ is
$-2z_2\,p^2$ which vanishes at $p^2=0$, 
it is natural to use
only $\Pi_\parallel^t$ and $\Pi_\parallel^s$ to define the extensions
of the parameter $a$.
With including these two parts only, the temporal and
the spatial components of 
$\Gamma_{\gamma\pi\pi}^\mu$ are given by
\begin{eqnarray}
 \Gamma_{\gamma\pi\pi}^0(0;q,-q)
 &=& 
  \frac{2q_0}{\widetilde{F}^2(\bar{q};T)}
   \Bigl[ \Pi_\perp^t(q_0,\bar{q};T)-
                \frac{1}{2}\Pi_\parallel^t(0,0;T)
         \Bigr], \nonumber\\
 \Gamma_{\gamma\pi\pi}^i(0;q,-q)
 &=& 
  \frac{-2q_i}{\widetilde{F}^2(\bar{q};T)}
    \Bigl[ \Pi_\perp^s(q_0,\bar{q};T)-
                \frac{1}{2}\Pi_\parallel^s(0,0;T)
         \Bigr].
\end{eqnarray}
Thus we define $a^t(T)$ and $a^s(T)$ as
\begin{eqnarray}
 a^t(\bar{q};T)
 = \frac{\Pi_\parallel^t(0,0;T)}
         {\Pi_\perp^t(q_0,\bar{q};T)} \ , 
 \quad
 a^s(\bar{q};T)
 = \frac{\Pi_\parallel^s(0,0;T)}
         {\Pi_\perp^s(q_0,\bar{q};T)} \ .
\end{eqnarray}
Here we should stress again that the pion momentum $q_\mu$ is
on mass-shell: $q_0 = v_\pi(\bar{q}) \bar{q}$.

In the HLS at one-loop level
the above
$a^t(\bar{q};T)$ and $a^s(\bar{q};T)$
are expressed as~\cite{HS:VD}
\begin{eqnarray}
 a^t (\bar{q};T)
 &=& a(0)\Biggl[ 1 + \frac{\bar{\Pi}_\parallel^t (0,0;T) -
     a(0)\bar{\Pi}_\perp^t (\bar{q},\bar{q};T)}
     {a(0) F_\pi^2(0;T)} \Biggr], \label{at expression}
\\
 a^s (\bar{q};T)
 &=& a(0)\Biggl[ 1 + \frac{\bar{\Pi}_\parallel^s (0,0;T) -
     a(0)\bar{\Pi}_\perp^s (\bar{q},\bar{q};T)}
     {a(0) F_\pi^2(0;T)} \Biggr], \label{as expression}
\end{eqnarray}
where $a(0)$ is defined in Eq.~(\ref{eq:Defa}) and 
$\Pi_\perp^{t,s}$ and $\Pi_{\parallel}^{t,s}$ express
the hadronic thermal corrections.

Before going to the analysis near the critical temperature,
let us study the temperature dependence of the parameters
$a^t(\bar{q};T)$ and $a^s(\bar{q};T)$ in the low temperature
region.
At low temperature $T \ll M_\rho$,
the hadronic thermal correction is dominated by the
contribution from thermal pions, and
$a^t$ and $a^s$ are expressed as~\cite{HS:VD}
\begin{equation}
 a^t \simeq a^s \simeq
         a(0) \left[
         1 + \frac{N_f}{12} \left( 1 - \frac{a^2}{4 a(0)} \right)
         \frac{T^2}{F_\pi^2(0;T)} \right],
  \label{at as form low T}
 \end{equation}
where $a$ is the parameter renormalized at the scale $\mu=M_\rho$,
while $a(0)$ is defined in Eq.~(\ref{eq:Defa}).
We think that the intrinsic temperature dependences are small in the
low temperature region, so that we use the values of parameters at
$T=0$ to estimate the temperature dependent correction to the above
parameters.
By using $F_\pi(0)=86.4\,\mbox{MeV}$,
$a(0) \simeq 2.31$ given in Eq.~(\ref{a0 val}) and
$a(M_\rho) = 1.38$ 
obtained through the Wilsonian matching for
$(\Lambda_{\rm QCD}\,,\,\Lambda) = (0.4\,,\,1.1)\,\mbox{GeV}$
and $N_f = 3$~\cite{HYc},
$a^t$ and $a^s$ in Eq.~(\ref{at as form low T}) are evaluated as
\begin{equation}
a^t \simeq a^s \simeq
a(0) \left[
  1 + 0.066 \left( \frac{T}{50\,\mbox{MeV}} \right)^2 
\right]
\ .
\end{equation}
This implies that the parameters $a^t$ and $a^s$ increase with
temperature in the low temperature region.
However, since the correction is small, we conclude that the
vector dominance is well satisfied in the low temperature region.

At higher temperature
the intrinsic thermal effects are important.
As we have shown in section~\ref{sec:VMT},
the parameters $(g,a)$ approach $(0,1)$
for $T \to T_c$ by the intrinsic temperature dependences,
and then the parametric vector meson mass $M_\rho$ vanishes.
Near the critical temperature
$\Pi_\perp^t$ and $\Pi_\perp^s$
in Eqs.~(\ref{at expression}) and (\ref{as expression})
approach the following expressions~\cite{HS:VD}:
\begin{eqnarray}
 \bar{\Pi}_\perp^t (\bar{q},\bar{q};T)
  \stackrel{T \to T_c}{\to} {}- \frac{N_f}{24}T^2 , \nonumber\\
 \bar{\Pi}_\perp^s (\bar{q},\bar{q};T)
  \stackrel{T \to T_c}{\to} {}- \frac{N_f}{24}T^2 .
\label{VM limits Pit Pis}
\end{eqnarray}
On the other hand, the functions
$\Pi_\parallel^{t}$ and
$\Pi_\parallel^{s}$ 
at the limit of $M_\rho/T \rightarrow0$ and $a\rightarrow1$
become
\begin{equation}
 \bar{\Pi }^{t}_\parallel(0,0;T)
 = \bar{\Pi }^{s}_\parallel(0,0;T)
 \rightarrow - \frac{N_f}{2} \tilde{I}_{2}(T) 
 = - \frac{N_f}{24} T^2 \ .
\label{VM limit PiVts}
\end{equation}
Furthermore, from Eq.~(\ref{a0 expression}), the parameter
$a(0)$ approaches $1$ for $M_\rho\rightarrow0$ and
$F_\sigma^2(M_\rho)/F_\pi^2(0)\rightarrow1$:
\begin{equation}
a(0) \rightarrow 1 \ .
\label{VM limit a0}
\end{equation}
{}From the above limits in Eqs.~(\ref{VM limits Pit Pis}),
(\ref{VM limit PiVts})
and (\ref{VM limit a0}), the numerators of 
$a^t(\bar{q};T)$ and $a^s(\bar{q};T)$ 
in Eqs.~(\ref{at expression}) and (\ref{as expression})
behave as
\begin{eqnarray}
 \bar{\Pi}_\parallel^t (0,0;T) - a(0)\bar{\Pi}_\perp^t
 (\bar{q},\bar{q};T) \to 0, \nonumber\\
 \bar{\Pi}_\parallel^s (0,0;T) - a(0)\bar{\Pi}_\perp^s
 (\bar{q},\bar{q};T) \to 0.
\end{eqnarray}
Thus 
we obtain
 \begin{equation}
  a^t(\bar{q};T), a^s(\bar{q};T) 
   \stackrel{T \to T_c}{\to} 1 \ .
 \end{equation}
This implies that the vector dominance is largely
violated near the critical temperature.

\section{S\lowercase{ummary}}
\label{sec:Sum}

In this write-up
we summarized main results which obtained in 
Refs.~\cite{HS:VMT,HS:VD}.

In section~\ref{sec:VMT}
we showed how the VM is realized in hot QCD.
It should be stressed that
the VM conditions in hot matter [Eq.~(\ref{g a z12:VMT})]
realized by the {\it intrinsic thermal effects}
are essential for the VM to take place in hot matter.

In section~\ref{sec:PaVVD}, 
we summarized main points to obtain a new prediction of the VM
made in Ref.~\cite{HS:VD}
on
the validity of vector dominance (VD) in hot matter.
In the HLS at zero temperature, 
the Wilsonian matching predicts $a \simeq2$~\cite{HYa,HYc}
which guarantees the VD of the
electromagnetic 
form factor of the pion.
Even at non-zero temperature,
this is valid as long as we consider the thermal effects
in the low temperature region
where the intrinsic temperature dependences are negligible.
We showed that, as a consequence of including the intrinsic effect,
the VD is largely violated at the critical temperature:
\begin{eqnarray*}
 a^t(\bar{p};T) \stackrel{T \to T_c}{\to} 1, \qquad
 a^s(\bar{p};T) \stackrel{T \to T_c}{\to} 1.
\end{eqnarray*}
In general, full temperature dependences include both hadronic and
intrinsic thermal effects.
Then there exists the violations of VD at generic temperature,
although at low temperature the VD is approximately satisfied.

In several analyses such as the one on the dilepton spectra 
in hot matter
done
in Ref~\cite{Rapp-Wambach:00}, the VD is assumed to be held even in
the high temperature region.
We should note that the analysis in Ref.~\cite{Pisarski} shows
that, if the VD holds, the thermal vector meson mass goes up.
Then the assumption of the VD, from the beginning, 
seems to exclude the possibility
of the dropping mass of the vector meson 
such as the one predicted by the 
Brown-Rho scaling~\cite{BR}.
Our result, which is consistent with the result in
Ref.~\cite{Pisarski} in some sense, indicates that the assumption of 
the VD may need to be weakened, at least in some amounts,
for consistently including the
effect of the dropping mass of the vector meson into the analysis.


\begin{theacknowledgments}
We would like to thank the organisers to give an opportunity
to present the talk at this happy occasion.
We are grateful to 
Doctor Youngman Kim, Professor Mannque Rho 
and Professor Koichi Yamawaki for useful discussions and comments
in writing Refs.~\cite{HS:VMT,HS:VD}.
MH would like to thank Professor Joe Schechter and Professor
Amir Fariborz for their hospitality during the stay in the
conference.

\end{theacknowledgments}


\end{document}